\documentclass[twoside,12pt]{article}
\usepackage{epsfig}
\usepackage{amsmath}
\usepackage{amssymb}

\usepackage{graphicx}  

\newcommand{\be}{\begin{equation}}
\newcommand{\ee}{\end{equation}}
\newcommand{\bea}{\begin{eqnarray}}
\newcommand{\eea}{\end{eqnarray}}

\begin{document}

\title{ \vspace{1cm} Emergent Gauge Symmetries and Particle Physics}
\author{Steven D.\ Bass $^{1,2}$ \\
\\
$^1$
Kitzb\"uhel Centre for Physics, Hinterstadt 12, 
A 6370 Kitzb\"uhel, Austria
\\
$^2$
Marian Smoluchowski Institute of Physics, Jagiellonian University, \\
PL 30-348 Krakow, Poland}
\maketitle
\begin{abstract} 
Hadron properties and interactions are emergent from QCD. 
Atomic and condensed matter physics are emergent from QED.
Could the local gauge symmetries of particle physics also 
be emergent? 
We give an introduction to this question and
recent ideas connecting it to the 
(meta)stability of the Standard Model Higgs vacuum.
With an emergent Standard Model 
the gauge symmetries would ``dissolve'' in the ultraviolet.
This scenario differs from unification models which exhibit
maximum symmetry in the extreme ultraviolet.
With emergence, new global symmetry violations would appear 
in higher dimensional operators.
\end{abstract}

\tableofcontents

\newpage

\section{Emergent particle physics}

One of the big surprises from the LHC is that the Standard Model works so well!
The Standard Model including QCD describes particle physics 
up to at least a few TeV as revealed in experiments at the
Large Hadron Collider and in low-energy precision experiments
such as electron electric dipole measurements and 
precision measurements of the fine structure constant 
$\alpha$.
Quantum Chromodynamics, QCD, gives us hadrons 
with their properties and interactions emergent 
from more fundamental quark and gluon degrees of freedom. 
The world of everyday experience 
(atoms, molecules, superconductors ...) 
is emergent from Quantum Electrodynamics, QED. 
The sun and nuclear reactors are powered by
radioactive $\beta$-decays through the weak interaction.

In high-energy particle physics the Higgs boson discovered at CERN in 2012 \cite{Aad:2012tfa,Chatrchyan:2012xdj}
completes the particle spectrum of the Standard Model.
The discovered boson behaves very Standard Model like.
It provides masses to the Standard Model particles.
With the masses and couplings measured at the LHC
the Standard Model works as a consistent theory up to the 
Planck scale with a Higgs vacuum very close to the border of
stable and metastable,
which may be a signal for new critical phenomena in the ultraviolet.
So far there is no evidence for new particles and interactions
either from collider experiments or from precision measurements.
Perhaps the symmetries of the Standard Model are more special
than previously anticipated.
Where do the gauge symmetries come from?
Do they unify in the ultraviolet or might the gauge symmetries of
the Standard Model be emergent in the infrared, ``dissolving'' 
in the ultraviolet close to the Planck scale?

In this article we discuss possible emergent gauge symmetries 
in particle physics. 
With main focus on phenomenology, 
we emphasise signatures that might show up in future experiments.

Emergence in physics occurs when a many-body system exhibits collective behaviour in the infrared
that is qualitatively different from that of its more primordial constituents as probed in the ultraviolet
\cite{Anderson:1972pca,Kivelson:2016}.
As an everyday example of emergent symmetry, consider a carpet which looks flat and translational invariant when looked at 
from a distance.
Up close, e.g. as perceived by an ant crawling on it, 
the carpet has structure and this translational invariance 
is lost.
The symmetry perceived in the infrared, 
e.g. by someone looking at it from a distance, 
``dissolves'' in the ultraviolet when the carpet is observed
close up.

New local gauge symmetries, where we make symmetry instead of breaking it,
are emergent in many-body quantum systems 
beyond the underlying QED symmetry and atomic interactions
\cite{Wen:2004ym,Levin:2004js,Zaanen:2011hm}.
Examples include
high temperature superconductors \cite{Baskaran:1987my,Sachdev:2015slk},
the Quantum Hall Effect \cite{Wilczek:1984dh}
and the A-phase of low temperature $^3$He \cite{Volovik:2003fe}.
Emergent Lorentz invariance is also observed in the infrared
limit of many-body quantum systems starting from a 
non-relativistic Hamiltonian, though some fine tuning 
may be needed to ensure the same effective limiting velocity 
$c$ for all species of (quasi-)particles~\cite{Levin:2005vf}.

To understand how emergent symmetry might work in particle
physics,
consider a statistical system near its critical point.
The long range tail is a 
renormalisable Euclidean quantum field theory with
properties described by the renormalisation group
\cite{Wilson:1973jj,Jegerlehner:1974dd,Jegerlehner:2013cta,Peskin:1995ev}.
The Landau-Ginzburg criterion tells us that 
fluctuations become important
for space-time dimensions of four or less.
This coincides with the dimensionality of space-time.
With four space-time
dimensions one finds an interacting quantum field
theory. 
With five or more space-time dimensions the physics 
reduces to a free field theory with long range modes decoupled.
With analyticity the Wick rotation means 
that the Euclidean quantum field theory is 
mathematically equivalent to the theory in Minkowski space.
Renormalisable theories with vector fields satisfy local 
gauge invariance~\cite{tHooft:1979hnm}.
If the physics contains massive gauge bosons,
then renormalisability 
\cite{tHooft:1971qjg,tHooft:1972tcz,Veltman:1968ki}
and 
unitarity 
\cite{LlewellynSmith:1973yud,Bell:1973ex,Cornwall:1973tb,Cornwall:1974km}
require Yang-Mills structure for the gauge fields 
(when we go beyond massive QED) together with a Higgs boson.

For emergence,
the key idea is that for a critical statistical system close 
to the Planck scale, the only long range correlations 
-- light mass particles --
that might exist in the infrared self-organise into multiplets just as they do in the Standard Model.
The vector modes would be the gauge bosons of U(1), SU(2) and SU(3).
In the self-organisation process
small gauge groups will most likely be preferred.
Gauge invariance is then exact 
(modulo spontaneous symmetry breaking)
in the energy domain of the infrared effective theory
with gauge invariance determining 
the number of polarisation degrees of freedom for the vector fields.
With parity violating chiral gauge interactions,
chiral anomaly cancellation in the ultraviolet limit of the effective theory then groups the fermions into families.
In this scenario the gauge theories of particle physics 
(and perhaps also General Relativity)
would be effective theories with characteristic energy of 
order the Planck scale~\cite{Weinberg:2009ca}.

The physics of the critical system residing close to the Planck
scale would be inaccessible to our experiments.
Theoretically, the challenge would be to understand the universality class of systems which exhibit identical 
critical behaviour, 
e.g. Standard Model like long-range behaviour.
By analogy, the physics of the extreme ultraviolet would be 
like probing the transition from quarks to hadrons as we go through the ultraviolet phase transition 
to (very possibly) completely different physics with different
degrees of freedom.
Whether the Standard Model is the unique stable low-energy limit
of the critical Planck system is an interesting subject for conjecture.

With emergence the Standard Model becomes an effective theory.
The usual Standard Model action is described by terms of 
mass dimension four or less.
In addition, 
with emergence one also finds an infinite tower of higher 
mass dimensional interaction terms with contributions 
suppressed by powers of a large ultraviolet scale 
which characterises the limit of the effective theory.
If we truncate the theory to include only operator terms 
with mass dimension at most four, then gauge invariant renormalisable interactions strongly constrain 
the global symmetries of the theory which are then inbuilt.
For example, electric charge is conserved and there is 
no term which violates lepton or baryon number conservation.
The dimension-four action describes long distance particle
interactions.
Going beyond mass-dimension four one finds gauge invariant
but non-renormalisable terms
where global symmetries are more relaxed and
which are suppressed by powers of 
the large ultraviolet scale associated with emergence.
Possible lepton number violation, 
also associated with Majorana neutrino masses, 
can enter at mass-dimension five, 
suppressed by a single power of the large emergence scale
\cite{Weinberg:1979sa}.
Baryon number violation can enter at dimension six, 
suppressed by the large emergence scale squared
\cite{Weinberg:1979sa,Wilczek:1979hc}.
The strong CP puzzle 
-- the absence of CP violation induced
by the non-perturbative glue which generates the large $\eta'$
meson mass --
might be connected to a possible new axion particle,
which is a postulated new pseudoscalar with coupling that enters at mass-dimension five
\cite{Weinberg:1977ma,Wilczek:1977pj}.
Dark matter \cite{Baudis:2018bvr}
might also involve dimension five (or higher) interactions
--
that is, with non-gravitational interaction strength 
very much suppressed by power(s) of the large emergence scale.
With the preference for small gauge groups, 
extra massive U(1)
gauge bosons might also be possible at mass-dimension four.

Possible emergent gauge symmetries in particle physics were discussed in early work by 
Bjorken \cite{Bjorken:2001pe,Bjorken:2010qx,Bjorken:1963vg},
Jegerlehner \cite{Jegerlehner:2013cta,Jegerlehner:1978nk,Jegerlehner:1998kt}
and Nielsen and collaborators \cite{Forster:1980dg}.
The key idea has enjoyed some recent renaissance, 
see the Perspectives article by Witten \cite{Witten:2017hdv}
as well as 
Refs.~\cite{Bass:2017nml,Jegerlehner:2018zxm,Giudice:2017pzm,tHooft:2007nis,Wetterich:2016qee}.
This article serves as an invitation to explore this physics 
and its phenomenology.

Gauge symmetries act on internal degrees of freedom
whereas global symmetries act on the Hilbert space.
That is, 
local gauge symmetries are 
properties of the description of a system and
global symmetries are properties of the system itself.

In the emergence scenario global symmetries would 
be restored with increasing energy 
(with energy exceeding symmetry breaking mass terms)
until we reach 
some very high energy where higher dimensional terms become important.
Then the system becomes increasingly chaotic
with possible lepton and baryon number violation and 
also possible Lorentz invariance violation in the extreme ultraviolet.
Additional sources of CP violation might also be possible
and important for understanding the matter-antimatter asymmetry generated in the early Universe.
This scenario differs from the situation in unification models
which exhibit maximum symmetry in the extreme ultraviolet and
where symmetries are spontaneously broken in the infrared, 
e.g. through coupling
to the Higgs and dynamical chiral symmetry breaking in QCD.

Lorentz invariance might also be emergent in the infrared
along with gauge symmetry.
Nielsen and collaborators considered the effect of 
adding a Lorentz violating term and found that it vanishes 
in the infrared through renormalisation group evolution, 
e.g. with Lorentz invariance 
emerging as an infrared fixed point \cite{Chadha:1982qq,Chkareuli:2001xe}.
In early work Bjorken suggested that the photon might 
be a Goldstone boson associated with spontaneous breaking of 
Lorentz invariance~\cite{Bjorken:1963vg}.
He also suggested that any violation of Lorentz invariance 
might be proportional to the ratio of 
the tiny cosmological constant scale, 0.002 eV, 
to the scale of emergence,
that is a ratio of about $10^{-27}$ and much beyond 
the range of present experiments~\cite{Bjorken:2001pe}.
Lorentz invariance is very strongly constrained by experiments~\cite{Kostelecky:2008ts,Ahlers:2018mkf}.
Being linked to Lorentz invariance, 
any violations of CPT symmetry would also be very tiny in this scenario.

The plan of the paper is as follows.
In Section 2 we review the status of the 
successful phenomenology of the Standard Model.
In Section 3 we discuss the role of gauge invariance in
defining the interactions of QED, QCD and the electroweak
Standard Model
and the constraints on global symmetries in these theories.
Following this introductory material,
Section 4 discusses the issue of vacuum stability and the
interplay of Standard Model parameters and the physics of
the extreme ultraviolet.
In Section 5 we compare the unification and emergence scenarios.
Section 6 discusses the Standard Model as a low energy expansion
and the physics of higher dimensional operators.
Here we discuss the signatures of emergent gauge symmetry that
might show up in future experiments.
Section 7 summarises the discussion of emergence 
versus unification and the open puzzles in this approach.

\section{Particle physics today - the very successful Standard Model}

The Standard Model is built on 
the gauge group of SU(3) colour, chiral SU(2) and U(1)
which are associated with QCD, weak interactions and QED 
respectively.
The ground state of 
QED is in the Coulomb phase, 
QCD in the confining phase and
weak interactions in a Higgs phase.
Whereas 
QED through massless photons has infinite range,
quarks and gluons in QCD can propagate up to about 1~fm before strong confinement forces take over and the relevant degrees of
freedom are colour singlet hadrons.
Weak interactions operate over a distance scale of about 0.01~fm
through massive W and Z boson exchanges.

The QCD and the SU(2) weak couplings are asymptotically 
free meaning that they decrease logarithmically in the ultraviolet.
In contrast, 
the QED coupling or fine structure constant increases
logarithmically in the ultraviolet with any divergence very much
above the Planck scale, the scale where quantum gravity effects 
are believed to become important.
Gluon-gluon interactions between gauge bosons in QCD and 
W-Z coupling with weak interactions give us asymptotic freedom.
The large QCD coupling in the infrared 
leads to confinement 
and dynamical chiral symmetry breaking
with pions and kaons as the corresponding Goldstone bosons.

In QED and QCD the photons and gluons interact with equal 
strength with left- and right-handed fermions.
The weak SU(2) interaction breaks parity and acts just on
left handed quarks and leptons grouped into lepton doublets
consisting of a charged lepton and neutrino and quark doublets
with an up-type quark (electric charge $+\frac{2}{3}$) and a 
down-type quark (electric charge $-\frac{1}{3}$).

The Standard Model has 18 parameters, 
or up to 27 
if we also include neutrino mixing and tiny neutrino masses: 
\begin{itemize}
\item
3 gauge couplings,
\item
15 in Higgs sector
(6 quark masses, 3 charged leptons, 4 quark mixing angles),
W and Higgs mass,
\item
9 neutrino parameters
(3 masses plus 6 mixing angles with Majorana neutrinos)
which might be connected to a dimension 5 operator.
\end{itemize}

The fermion multiplet structure is reproduced three times
in families, also called generations.
Precision measurements of Z$^0$ decays at the LEP experiments
at CERN revealed 
the number of light neutrinos as 
$2.984 \pm 0.008$ \cite{pdg:2018}.
The number of neutrino families is 
determined independently
from the cosmic microwave background.
One finds $3.13 \pm 0.32$ \cite{Aghanim:2018eyx}.

Photons and gluons are massless.
The masses of W and Z bosons and of charged fermions 
(leptons and quarks) come from the Higgs sector.
The W and Z bosons have mass 80 and 91 GeV and the Higgs boson has mass 125 GeV.
The charged leptons and their masses are
\begin{equation}
m_e = 0.51 \ {\rm MeV}, \ \ \ \ \
m_{\mu} = 105.66 \ {\rm MeV}, \ \ \ \ \
m_{\tau} = 1776.86 \pm 0.12 \ {\rm MeV} .
\end{equation}
Neutrinos come with tiny masses, 
see Section 6.1 below, as evidenced by neutrino oscillation data \cite{Balantekin:2018azf}.
The down-type quarks have masses
\begin{equation}
m_d = 5 \ {\rm MeV}, \ \ \ \ \
m_s = 93 \ {\rm MeV}, \ \ \ \ \
m_b = 4.18 \ {\rm GeV}
\end{equation}
whereas the up-type quarks have masses
\begin{equation}
m_u = 2 \ {\rm MeV}, \ \ \ \ \
m_c = 1.27 \ {\rm GeV}, \ \ \ \ \
m_t = 173.1 \pm 0.9 \ {\rm GeV} .
\end{equation}
Strong interactions of QCD are an additional source of mass
generation.
Scalar confinement of 
quarks and gluons in the proton generates the large proton mass.
About 99\% of the mass of the 
hydrogen atom 938.8 MeV
is associated with the confinement potential
with the masses of 
the electron 0.5 MeV and
the proton 938.3 MeV.
Inside the proton 
the masses of the proton's constituent two up quarks and 
one down quark contribute about 9 MeV.

The number of degrees of freedom for the gauge bosons 
are dependent on the ground state, whether in the Coulomb,
confining or Higgs phase.
The massless photons 
carry two transverse polarisation states
whereas
the massive W and Z bosons have three polarisations 
with longitudinal polarisation also included.
Massless gluons come with two transverse polarisations
but are always virtual because of confinement meaning
that longitudinal gluons can play a role.
The charge-neutral scalar Higgs has just one degree of freedom.

Particle physics experiments measure interactions described
by the Standard Model action with mass dimension at most four.
Higher dimensional terms are suppressed 
by powers of some large ultraviolet scale which we are not
yet sensitive to with the present energies and precision of
our experiments.
The Lagrangian restricted to terms with mass dimension at most 
four is renormalisable.
Ultraviolet divergences from loop diagrams can be 
self-consistently absorbed in a re-definition of the parameters
\cite{Bjorken:1965zz}.
Renormalisability tells us that, 
with parity violating vector interactions,
we have to worry about chiral anomaly cancellation in the
ultraviolet.
For example, the triangle Feynman diagrams with 
two photon or gluon vertices and a chiral Z$^0$ coupling 
cannot be gauge invariantly renormalised unless the 
relevant charges of the fermions propagating in the loop 
sum to zero.
This in turn groups the fermions into families with
combinations of charges perfectly aligned to cancel
any local chiral anomalies, meaning that gauge invariance
and renormalisability are preserved.

Three families are also needed for CP violation with the 
Cabibbo-Kobayashi-Maskawa (CKM) matrix.
This is a unitary $3 \times 3$ quark mixing matrix
with three real mixing parameters and one CP violating angle.
The measured mixing parameters from the Cabibbo angle 
to heavy-quark mixing angles are consistent with a 
unitary matrix and no new CP violating physics 
in the energy range of present experiments, e.g. up to LHC
energies.
Precision studies of the electroweak Standard Model come 
from measurements at the Z$^0$ pole from LEP at CERN and SLD 
at SLAC -- for a review see \cite{Altarelli:2013tya}.
Key observables include the weak mixing angle 
$\sin^2 \theta_W$,
forward-backward asymmetries and $\tau$ lepton polarisation.

Our everyday experience is described by just the first 
generation of light quarks and leptons 
(protons, neutrons, pions, electrons, plus neutrinos from 
 the Sun).
However, our existence is not insensitive to the physics of 
the highest scales.
After radiative corrections 
the values of the top-quark and Higgs boson masses 
are essential for whether 
the electroweak particle physics vacuum is stable or not
\cite{Degrassi:2012ry,Buttazzo:2013uya,Bezrukov:2012sa,Bezrukov:2014ina,Alekhin:2012py,Masina:2012tz,Hamada:2012bp,Bednyakov:2015sca}.
With the value of the top quark mass measured at the LHC, 
the Higgs mass is about the minimum necessary for the 
Standard Model vacuum to be stable, see Section 4 below.
One finds a delicate balance between Standard Model 
parameters and the physics of the extreme ultraviolet.

Going beyond the Standard Model,
there were hopes of the Standard Model couplings meeting
in the ultraviolet,
perhaps with help from some new particles or interactions.
This would lead to unification of the three Standard Model 
forces,
generalising the electroweak unification of electromagnetism
and weak interactions to include QCD.
With just Standard Model couplings, 
they nearly do meet but not exactly -- see Section 4 below.

The observed matter antimatter asymmetry in the Universe 
requires some extra source of CP violation beyond
the quark mixing described by the 
CKM matrix in the electroweak Standard Model. 
In the neutrino sector
recent measurements by the T2K Collaboration 
in Japan are consistent with CP violation at
the level of two standard deviations \cite{Abe:2018wpn,Abe:2019vii}.
In addition,
low-energy precision experiments are used
to look for possible new sources of CP violation.
Key experiments involve the search for electric dipole
moments
\cite{Andreev:2018ayy,Afach:2015sja,Chupp:2014gka}
plus precision measurements of CP sensitive observables 
in positronium decays
\cite{Bass:2019ibo,Moskal:2016moj,Bernreuther:1988tt}.
So far there is no evidence for new extra sources of CP violation from these low-energy precision experiments.

\subsection{Precision QED tests}

QED is the most accurately tested theory with 
remarkable precision achieved in different measurements of 
the fine structure constant $\alpha$.
The most accurate determinations of $\alpha$ come from precision measurements of 
the electron's anomalous magnetic moment~\cite{Hanneke:2008tm}
and atom interferometry measurements with 
Caesium, Cs \cite{Parker:2018sc}.
The electron anomalous magnetic moment 
$a_e = (g-2)/2$
is generated by radiative corrections,
which have been evaluated to tenth-order
in QED perturbation theory
plus tiny QCD and weak contributions~\cite{Aoyama:2017uqe}.
The electron $a_e$ value
gives a precision measurement of $\alpha$ 
(modulo any radiative corrections from 
 new physics beyond the Standard Model).
Atom interferometry experiments with Cs provide a more direct determination
(less sensitive to details of radiative corrections)
but also involve a combination of parameters measured 
in experiments:
the Rydberg constant $R_{\infty}$,
the ratio of the atom to electron mass $m_{\rm atom}/m_e$
and new precision measurements of 
the Cs mass from recoil of a Cs atom in an atomic lattice,
viz.
$
\alpha^2
=
(2 R_{\infty} / c)
\ ( m_{\rm atom} / m_e) \ ( h / m_{\rm atom} ).
$
(Here $c$ is the speed of light and $h$ is Planck's constant.)
Comparing these different determinations of $\alpha$ 
gives a precision test of QED
as well as constraining possible new physics scenarios.
Any ``beyond the Standard Model'' effects involving 
new particles active in radiative corrections will 
enter $a_e$ but not the Cs measurements.
The new most accurate Cs atomic physics measurement
corresponds to 
\begin{equation}
a_e^{\rm exp} - a_e^{\rm th}|_{\rm Cs}
=
(-88 \pm 36 ) \times 10^{-14}
\end{equation}
when we substitute the $\alpha$ value measured in these
atomic physics experiments into the perturbative expansion
of $a_e$
to obtain the value $a_e^{\rm th}|_{\rm Cs}$.
That is, one finds agreement to 1 part in $10^{12}$.

\subsection{QCD and emergent hadrons}

QCD is fundamentally different because of confinement
in the infrared.
Quarks carry a colour charge and interact through coloured
gluon exchange, like electrons interacting through photon
exchange in QED.
QCD differs from QED in that
gluons also carry colour 
charge whereas photons are electrically neutral.
This means that the Feynman diagrams for QCD include
3 gluon and 4 gluon vertices (as well as the quark gluon vertices) and that gluons self-interact.
The three gluon vertex leads to gluon bremsstrahlung 
resulting in gluon induced jets of hadronic particles
which were first discovered in high energy $e^- e^+$ collisions at DESY.
The decay amplitude for $\pi^0 \to 2 \gamma$ and 
the ratio of cross-sections
for hadron to muon-pair production in 
high energy electron-positron collisions, $R_{e^- e^+}$,  
are each proportional to the number of dynamical colours $N_c$,
giving an experimental confirmation of $N_c =3$.

In the infrared quark-gluon interactions become strong.
Low energy QCD is characterised by confinement and dynamical
chiral symmetry breaking.
The physical degrees of freedom are emergent hadrons 
(protons, mesons ...) 
as confined bound states of quarks and gluons. 
Spontaneous chiral symmetry breaking is associated with a 
non-vanishing chiral quark condensate. 
The light mass pions (and kaons) are 
the corresponding would-be Goldstone bosons
with mass squared proportional to the light quark masses,
$m_{\pi}^2 \sim m_q$.
In the isosinglet channel
non-perturbative gluon dynamics 
increase the masses of the $\eta$ and $\eta'$ mesons 
by about 300-400 MeV relative to the masses they would have 
if they were pure Goldstone states \cite{Bass:2018xmz}.

The proton's mass and spin are emergent from quark and gluon
degrees of freedom.

High energy deep inelastic scattering experiments probe the
deep structure of hadrons by scattering high energy electron
or muon beams off hadronic targets. 
Deeply virtual photon exchange acts like a microscope which
allows us to look deep inside the proton.
These experiments reveal a proton built of 
nearly free fermion constituents, called partons.
Quark and gluon partons play a vital role in high energy
hadronic collisions, 
e.g., at the LHC~\cite{Altarelli:2013tya}.
Deep inelastic scattering experiments also tell 
us that about 50\% of the proton's momentum 
perceived at high $Q^2$ is carried by gluons,
consistent with the 
QCD prediction for the deepest structure of the proton.
Polarised deep inelastic scattering experiments 
have taught us that just about 30\% of the proton's spin 
of one half
is carried by its quarks~\cite{Bass:2004xa}.
The rest is carried by gluons and by quark and 
gluon orbital angular momentum.
Confinement generates a transverse momentum scale in 
the proton leading to finite quark and gluon orbital 
angular momentum contributions.
Scalar confinement also induces
dynamical chiral symmetry breaking, 
e.g., in the Bag model
the Bag wall connects left and right handed quarks 
leading to quark-pion coupling and the pion cloud of the nucleon \cite{Thomas:1982kv}.
The pion cloud takes further orbital angular momentum
through quark-pion coupling in the nucleon~\cite{Bass:2009ed}.
One finds a consistent picture where pion cloud dynamics,
modest gluon polarisation 
(up to about 50\% of the proton's spin at the scale of 
 typical deep inelastic experiments) and perhaps non-local
gluon topology describe the internal spin structure of the
proton \cite{Bass:2004xa,Aidala:2012mv}.

Hadron physics is our first example of emergence in particle physics with change to totally new degrees of freedom as one
goes through the confinement transition from coloured quarks 
and gluons to colour-neutral hadrons.

\section{Global and local gauge symmetries}

We next focus on the symmetries in the particle physics Lagrangian, 
first with QED and then QCD and the electroweak Standard Model.
Our aim in this Section is to show the role that 
local gauge invariance plays in constraining the 
interaction terms and global symmetries of the Standard Model.
This discussion will lead into the exploration of 
global symmetry breaking terms with higher mass dimension in
Sections 5 and 6.

Local gauge symmetries determine the dynamics.
Poincare invariance is an important part of quantum field theory together with the associated discrete symmetries of 
P, C, T, CP (which can be broken) 
and fundamental CPT symmetry which is exact.
The usual particle physics Lagrangian includes fields 
and interaction terms with mass dimension at most four.
For example, 
the mass dimensions of fermion fields $\psi$,
scalar bosons $\phi$ and vector fields $A_{\mu}$ 
and their interaction terms are 
\begin{itemize}
\item
$[ \psi ] = \frac{3}{2}$
\item
$[ \phi ] = 1$
\item
$[ A_{\mu} ] = 1$
\item
$[ m ] = 1$
\item
$[ \partial_{\mu} ] = 1$
\item
$[ m \bar{\psi} \psi ] = 4$
\item
$[ \partial^{\mu} \phi \ \partial_{\mu} \phi ] = 4$
\end{itemize}

Starting with the theoretical Lagrangian, 
Noether's theorem tells us that
there are conserved currents associated with continuous global
symmetries.
For example, translational invariance is associated with 
momentum conservation.
Rotational invariance is associated with angular momentum
conservation.
Electric charge conservation is associated 
with global U(1) invariance in QED and the conserved 
(and gauge invariant) vector current.
Invariance under global axial rotations of the phase of
the fermion fields leads to the Noether current 
$j_{\mu 5} = {\bar \psi} \gamma_{\mu} \gamma_5 \psi$.
Conservation of $j_{\mu 5}$ 
which also corresponds 
to fermion helicity conservation
is softly broken by fermion mass terms
(and also sensitive to anomalous terms in its divergence equation
 in the singlet channel where one couples through two gauge-boson
 intermediate states~\cite{Adler:1969gk,Bell:1969ts}).

Fermion masses which break chiral symmetry between left- and
right-handed fermions represent a continuous deformation of 
the massless theory.
Gauge boson masses which enter through 
the Higgs mechanism
with spontaneous symmetry breaking 
change the degrees of freedom meaning we have a different 
theory
(where longitudinal polarisation of the gauge bosons becomes 
 physical).

If we truncate the theory to operators of mass dimension at most four, 
then the global symmetries are strongly constrained 
by the operators that are allowed by gauge invariance
and renormalisability.
Particle masses and global symmetry breaking becomes less important with increasing energy,
especially when the energy is much greater than the
particle masses, $E \gg m$.
Global symmetries which are compelled to hold at dimension four
can be broken in non-renormalisable higher dimensional operators
which are suppressed by powers of some large ultraviolet scale
and become active only in the extreme ultraviolet
\cite{Jegerlehner:2013cta,Witten:2017hdv}. 
Examples include lepton and baryon number violation discussed 
in Section 6.
If we allow for new higher dimensional terms in the action,
then we find increasing restoration of global symmetries 
with increasing energy and resolution
until we become sensitive to these higher dimensional terms,
say at energies within about 0.1\% of the large ultraviolet scale.

We next look in detail at gauge symmetry which is intrinsic 
to particle physics interactions.

\subsection{Quantum Electrodynamics}

Quantum Electrodynamics, QED, 
follows from requiring that the physics is 
invariant under local U(1) changes of the phase of charged
particles,
e.g. the electron,
viz.
\begin{equation}
\psi (x) \rightarrow e^{i \alpha (x)} \psi (x)
\end{equation}
where 
$\alpha (x)$ is a function of the space-time co-ordinates.
Derivative terms $\partial_{\mu}$ acting on
$\psi$ will also act on the phase factor $\alpha (x)$
so that the phase factor does not flow through the combination $\partial_{\mu} \psi$.
Instead, consider the gauge covariant derivative
\begin{equation}
\partial_{\mu} \mapsto D_{\mu} = \partial_{\mu} + i e A_{\mu}
\end{equation}
where $A_{\mu}$ 
is the gauge field and 
$e$ is the electric charge with the fine structure constant
\begin{equation}
\alpha = e^2 / 4 \pi .
\end{equation}
With $A_{\mu}$
transforming under the phase rotation in Eq.(5) as
\begin{equation}
A_{\mu} (x) \rightarrow A_{\mu}' (x) 
  = A_{\mu} (x) - {1 \over e} \partial_{\mu} \alpha (x) ,
\end{equation}
the combination $D_{\mu} \psi$ transforms as
\begin{equation}
D_{\mu} \psi \rightarrow e^{i \alpha(x)} D_{\mu} \psi
\end{equation}
and the phase factor is pulled through the derivative term.
The gauge field $A_{\mu}$ describes the photon after quantisation.
The photon field tensor 
$
F_{\mu \nu} = \partial_{\mu} A_{\nu} - \partial_{\nu} A_{\mu}
$
is invariant under transformations of $A_{\mu}$, Eq.~(8).

The QED Lagrangian
\begin{equation}
{\cal L} =
\bar{\psi} i \gamma^{\mu} D_{\mu} \psi 
- m {\bar \psi} \psi
- {1 \over 4} F_{\mu \nu} F^{\mu \nu} 
\end{equation}
is gauge invariant.
After quantisation it describes the QED dynamics:
the electron and photon propagators and the electron-photon
interaction vertex.
The first term includes the kinetic energy for the electron
together with the electron photon interaction, 
$e {\bar \psi} \gamma^{\mu} A_{\mu} \psi$.
The second term is the electron mass.
The third term describes the photon kinetic energy.

\begin{itemize}
\item
Real photons come with two transverse polarisations.
The time and longitudinal components of $A_{\mu}$ 
(for real photons)
are really not dynamical degrees of freedom.
They can be set equal to zero by a suitable choice of
gauge - the radiation or Coulomb gauge, $\nabla . A =0$.
In general, under a Lorentz transformation 
$A_{\mu}$ does not transform as a four-vector
but is supplemented by an additional gauge term
which ensures that only gauge invariant 
Maxwell equations are Lorentz covariant~\cite{Bjorken:1965zz}.

\item
Conserved electric charge corresponds to global U(1)
transformations.

\item
The electron mass term breaks the chiral symmetry between 
left- and right-handed electrons.
Helicity is conserved for massless electrons.
(Without the electron mass, left- and right-handed electron
 fields, 
 $\psi_L = \frac{1}{2}(1-\gamma_5) \psi$ and
 $\psi_R = \frac{1}{2}(1+\gamma_5) \psi$, 
 transform independently under chiral rotations.)

\item

Mass is important in quantum field theories and is needed
even in QED;
charged particles should also carry mass 
\cite{Gribov:1981jw,Morchio:1985re,Gomez:2019sxl}.
Starting from Eq.~(10)
one cannot perturbatively renormalise massless QED on-shell.
If one renormalises the massive theory on-shell and then takes
the mathematical limit that the electron mass goes to zero, 
then the Landau pole in the running coupling gets pulled towards the infrared,
\begin{equation}
\alpha (\lambda_{\rm UV}^2) =
\frac{\alpha (m^2)}
{1 - \frac{\alpha(m^2)}{3\pi} 
\ln \frac{\lambda_{\rm UV}^2}{m^2}} .
\end{equation}

\item

If we treat QED as an effective theory, then one might add also 
the gauge invariant but non-renormalisable Pauli term
\begin{equation}
i \frac{e}{2M} \
{\bar \psi} (\gamma^{\mu} \gamma^{\nu}
- \gamma^{\nu} \gamma^{\mu} ) \psi \ F_{\mu \nu}
\end{equation}
which is suppressed by power of some large mass scale $M$
\cite{Weinberg:1995mt}.
This is our first example of a 
non-renormalisable dimension 5 operator.
It gives a contribution to the electron magnetic moment of 
$4 e / M$.
Experiment through Eq.~(4) constrains
$M$ to be at least $3 \times 10^{10}$ GeV.

The dimension 5 Pauli term with finite large $M$ gives a
finite value of $\alpha$ without Landau pole
or triviality issues \cite{Djukanovic:2017thn}.

\end{itemize}

\subsection{QCD and non-abelian gauge theories}

The same arguments that led to QED can be generalised to 
non-abelian groups, e.g. SU(2) and SU(3).
For QCD we require the physics to be invariant under
\begin{equation}
\Psi (x) \rightarrow U(x) \Psi (x)
\end{equation}
where
\begin{equation}
U(x) = e^{i {1 \over 2} {\vec \lambda}.{\vec \alpha} (x)} 
\end{equation}
and $\lambda_a$ are the eight $3 \times 3$ Gell-Mann matrices.
Here $\Psi$ is a 3-dimensional spinor corresponding 
to quarks carrying the SU(3) colours red, green and blue.
The QCD gauge covariant derivative is
\begin{equation}
D_{\mu} \Psi =
\biggl[ \partial_{\mu} 
+ \frac{i}{2} g_3 \vec{\lambda}.\vec{A}_{\mu} 
 \biggr] \Psi
\end{equation}
where $g_3$ is the SU(3) colour charge and 
the gluon gauge fields
$A_{\mu} = \frac{1}{2} \vec{\lambda}.\vec{A}_{\mu}$ 
satisfy the transformation rule
\begin{equation}
A_{\mu} (x) \rightarrow A_{\mu} \ ' (x) 
=
U A_{\mu} U^{-1}  + \frac{i}{g_3} (\partial_{\mu} U) U^{-1} .
%
%
\end{equation}
Corresponding to SU(3) phase rotations (13) and (14),
the QCD gauge covariant derivative transforms as
\begin{equation}
D_{\mu} \rightarrow U D_{\mu} U^{\dagger} .
\end{equation}
There are eight gluon fields.
The field tensor for these gauge fields is
\begin{equation}
G_{\mu \nu}^a 
= \bigl[ D_{\mu}, D_{\nu} \bigr]_- 
=
\partial_{\mu} A_{\nu}^a - \partial_{\nu} A_{\mu}^a
- g_3 f_{abc} A_{\mu}^b A_{\nu}^c
\end{equation}
where the $f_{abc}$ are the structure constants of SU(3),
$[t^a, t^b] = i f_{abc} t^c$ 
with $t^a = \frac{1}{2} \lambda^a$.
Putting things together the QCD Lagrangian
\begin{equation}
{\cal L} =
\bar{\Psi} i \gamma^{\mu} D_{\mu} \Psi - m {\bar \Psi} \Psi
- {1 \over 4} {\rm Tr} \ G_{\mu \nu} G^{\mu \nu} 
\end{equation}
is invariant under the SU(3) gauge transformations
in Eqs.~(13) and (16).

\begin{itemize}

\item

The QCD running coupling is asymptotically free.  
At leading order
\begin{equation}
\alpha_s (K^2) =
\frac{g_3^2}{4 \pi} =
\frac{4 \pi}
{\beta_0 \ln ( K^2 / \Lambda_{\rm QCD}^2 )} .
\end{equation}
Here $K^2$ is the four-momentum transfer squared,
$\beta_0 = \frac{11}{3} N_c - \frac{2}{3} f$
where $N_c=3$ is the number of colours and $f$ is
the number of active flavours;
$\Lambda_{\rm QCD}$ is 
the renormalisation group invariant QCD infrared scale, 
about 200 MeV.

\item

Rising $\alpha_s$ in the infrared leads to QCD confinement of
quarks and gluons.
A scalar confinement potential
leads to dynamical mass generation through the strong interactions. 
It also spontaneously breaks the chiral symmetry between 
left- and right- handed quarks
with pions and kaons as the would-be Goldstone bosons
with mass squared
$m^2 \propto m_q$
where $m_q$ is the light-quark mass.
The light-quarks then acquire an effective large constituent 
quark mass of about 300 MeV in the infrared.

\item

In the isosinglet channel the $\eta$ and $\eta'$ mesons are
too heavy by about 300--400 MeV to be pure Goldstone bosons.
They receive extra mass from non-perturbative gluon processes 
in the flavour-singlet channel~\cite{Bass:2018xmz}
connected to non-perturbative gluon topology and the QCD axial anomaly
in the divergence of the flavour-singlet axial-vector current
\cite{Adler:1969gk,Bell:1969ts}.
While the non-singlet axial-vector currents like
$J_{\mu 5}^{(3)} = 
\bar{u} \gamma_{\mu} \gamma_5 u 
- \bar{d} \gamma_{\mu} \gamma_5 d $
are partially conserved 
(they have just mass terms in the divergence), 
the singlet current
$
J_{\mu 5} = \bar{u}\gamma_\mu\gamma_5u
+ \bar{d}\gamma_\mu\gamma_5d + \bar{s}\gamma_\mu\gamma_5s 
$
satisfies the divergence equation 
\begin{equation}
\partial^\mu J_{\mu5} = 6 Q_t
+ \sum_{k=1}^{3} 2 i m_k \bar{q}_k \gamma_5 q_k 
\end{equation}
where 
\begin{equation}
Q_t 
= {\alpha_s \over 8 \pi} G_{\mu \nu} {\tilde G}^{\mu \nu}
\end{equation}
is called the topological charge density.
Here $G_{\mu \nu}$ is the gluon field tensor and
${\tilde G}^{\mu \nu} = 
{1 \over 2} \epsilon^{\mu \nu \alpha \beta} G_{\alpha \beta}$.
Beyond perturbative QCD, $\int d^4 x Q_t$ 
quantises with integer or 
fractional values~\cite{Crewther:1978zz}.
Non-perturbative gluon processes act 
to connect left- and right-handed quarks 
whereas 
left- and right-handed massless quarks 
propagate independently in perturbative QCD
with helicity conserved for massless quarks.

The non-perturbative gluon dynamics which generate 
the large $\eta'$ mass also has the potential to induce 
strong CP violation in QCD.
Provided the quark masses are non-zero,
e.g. 
through the Yukawa couplings associated with 
the Higgs mechanism, 
one finds a new effective CP-odd term in the 
QCD Lagrangian at mass dimension-four \cite{Weinberg:1996kr}
\begin{equation}
{\cal L} = - \theta Q_t .
\end{equation}
Experimentally, $\theta < 10^{-10}$ \cite{Afach:2015sja}.
We return to this physics in Section 6.2 below.
Finite $\theta$ values in Eq.~(23) are induced 
by non-zero quark masses which enter at mass dimension four.
The suppression of strong CP violation may involve a 
new axion particle 
(which enters with mass and coupling at mass dimension five)
with delicate interplay of interactions at dimension four and dimension five. 

\end{itemize}

\subsection{The Higgs mechanism and the Standard Model}

Before discovery of the W and Z bosons parity violating weak interactions
were described using Fermi's four-fermion point interaction
with coupling
\begin{equation}
\frac{1}{\sqrt{2}} G_F = \frac{g^2}{8 m_W^2} .
\end{equation}
The coupling here $G_F$ comes with mass dimension minus-two
with the four-fermion interaction 
violating unitarity and renormalisability 
signalling the need for new physics at a deeper level
-- the Standard Model.

We now know in the Standard Model that weak interactions
are mediated by massive W and Z gauge boson exchange with
left-handed fermion doublets and gauge group SU(2).
Restricting to one 
family of fermions the lepton and quark doublets are
\begin{equation}
L =
\left(\begin{array}{c}
\nu 
\\
e
\end{array}\right) 
\ , \ \ \ \ \ \
Q =
\left(\begin{array}{c}
u 
\\
d* 
\end{array}\right)_c
\end{equation}
where $u$ is the up quark and $d$ the down quark.
Going beyond the first fermion family, $d*$ in the lower component of the quark doublet includes mixing from the 
Cabibbo angle (including strange quarks) and the full 
Cabbibo-Kobayashi-Maskawa matrix
taking into account the three families of fermions.
The subscript $c$ denotes the three colours of quarks 
(red, green and blue) associated with QCD.

QED and weak interactions unify through mixing between 
the charge neutral photon and Z boson
with the gauge group
SU(2)$_L \otimes$U(1).
We let $B_{\mu}$ denote 
the U(1) gauge boson and $W_{\mu}^i$ denote the SU(2) bosons.
The U(1) gauge bosons interact equally with left- and right-
handed fermions.
Fermions transform under the SU(2) and U(1) 
gauge transformations as
\begin{equation}
\Psi_L (x) \rightarrow 
e^{i {1 \over 2} {\vec \tau}.{\vec \alpha} (x)} \Psi_L (x)
\ \ \ \ \ {\rm and} \ \ \ \ \
\Psi (x) \rightarrow 
e^{i {y \over 2} \beta (x)} \Psi (x)
\end{equation}
with gauge covariant derivative
\begin{eqnarray}
D_{\mu} \Psi_L 
&=&
\biggl[ \partial_{\mu} 
+ \frac{1}{2} i g \vec{\tau}.\vec{W}_{\mu} 
+ \frac{1}{2} i g' y B_{\mu} \biggr] \Psi_L
\nonumber \\
D_{\mu} \Psi_R 
&=&
\biggl[ \partial_{\mu} + \frac{1}{2} i g' y B_{\mu} \biggr] \Psi_R .
\end{eqnarray}
Here 
$\tau$ denotes the SU(2) Pauli matrices and
$g$ and $g'$ are the SU(2) and U(1) couplings.
The electric charge is $ Q = t_3 + y/2 $
where $t_3 = \tau_3/2$. 
The hypercharge $y$ is 
$y=-1$ for left-handed leptons $l_L$, 
$y=-2$ for the right-handed leptons $l_R$,
$y=\frac{1}{3}$ for the left-handed quarks,
$y=-\frac{2}{3}$ for right-handed down-type quarks, 
and
$y=\frac{4}{3}$ for right-handed up-type quarks.
With these assignments the electron carries electric charge 
and the neutrino is electric charge neutral.

The electric charge neutral gauge bosons mix as
\begin{equation}
\left(\begin{array}{c}
W_{\mu}^3
\\
B_{\mu}
\end{array}\right)
=
\left(\begin{array}{cc}
\cos {\theta_W} & \sin {\theta_W}
\\
- \sin {\theta_W} &  \cos {\theta_W}
\end{array}\right)
\left(\begin{array}{c}
Z_{\mu} 
\\
A_{\mu}
\end{array}\right)
\end{equation}
where $A_{\mu}$ is the photon field, 
$Z_{\mu}$ is the Z boson field and 
$\theta_W$ is the Weinberg angle.
The neutral current with left-handed fermions 
interaction is then
\begin{equation}
i g \sin \theta_W 
{\bar \Psi}_L \gamma^{\mu} \
\biggl[
A_{\mu} (t_3 + \frac{1}{2} y)
+  Z_{\mu} (-\frac{1}{2} y \tan \theta_W I 
            + \cot \theta_W t_3)
\biggr] \Psi_L .
\end{equation}
The photon is massless provided that
\begin{equation}
g \sin \theta_W = g' \cos \theta_W = e
\end{equation}
or
\begin{equation}
\tan \theta_W = g' / g .
\end{equation}
Mixing fixes the Weinberg angle $\theta_W$
\begin{equation}
\cos \theta_W = \frac{g}{\sqrt{g^2 + g'^2}}, \ \ \
\sin \theta_W = \frac{g'}{\sqrt{g^2 + g'^2}}
\end{equation}
-- that is,
\begin{equation}
W_{\mu}^{\pm} 
= \frac{1}{\sqrt{2}} ( W_{\mu}^1 \mp W_{\mu}^2 ),
\ \ \
B_{\mu} 
= \frac{-g' Z_{\mu} + g A_{\mu}}{\sqrt{g^2 + g'^2}},
\ \ \ 
W_{\mu}^3 
= \frac{g Z_{\mu} + g' A_{\mu}}{\sqrt{g^2 + g'^2}} .
\end{equation}
The W$^{\pm}$ connect different members of the electroweak
lepton and quark doublets whereas the photon 
and Z$^0$ 
are electric-charge neutral bosons.

Naively, gauge boson mass terms break gauge invariance:
the mass term 
$m^2 W_{\mu} W^{\mu }$ is not invariant 
under gauge transformations of the $W_{\mu}^a$.
Gauge invariance is maintained through the Higgs mechanism
\cite{Higgs:1964ia,Higgs:1964pj,Higgs:1966ev,Englert:1964et}.
One adds the scalar doublet $\Phi$ with potential
\begin{equation}
V(\Phi) = \frac{1}{2} \mu^2 \Phi^2 + \frac{1}{4} \lambda \Phi^4
\end{equation}
and transforming as
\begin{equation}
\Phi (x) \rightarrow 
e^{i {1 \over 2} {\vec \tau}.{\vec \alpha} (x)} \Phi (x)
\ \ \ \ \ {\rm and} \ \ \ \ \
\Phi (x) \rightarrow 
e^{i \frac{y}{2} \beta (x)} \Phi (x)
\end{equation}
under the SU(2) and U(1) gauge transformations (26).
In Eq.~(34) $\lambda \geq 0$ to ensure that the potential 
has a finite minimum, as required for vacuum stability.
The Higgs scalar comes with the covariant derivative coupling
\begin{equation}
D_{\mu} \Phi =
\biggl[ \partial_{\mu} 
+ \frac{1}{2} i g \vec{\tau}.\vec{W}_{\mu} 
+ \frac{1}{2} i g' y_{\phi} B_{\mu} \biggr] \Phi .
\end{equation}
Here the scalar hypercharge 
$y_\phi = +1$ for the Standard Model
to ensure that the photon does not couple 
to the Higgs boson and $\phi^+$ has the correct charge.
For $\mu^2>0$ Eq.~(34) describes the potential 
for a particle with mass $\mu$.
More interesting is the case $\mu^2 < 0$. 
In this case the potential has a minimum at
\begin{equation}
| \Phi | 
= {v \over \sqrt{2}} \equiv \sqrt{- {\mu^2 \over 2 \lambda}}
\end{equation}
where $v$ is the vacuum expectation value, vev.
One takes the vev to be real,
\begin{equation}
\langle \Phi \rangle =
\frac{1}{\sqrt{2}}
\left(\begin{array}{c}
0 
\\
v 
\end{array}\right) ,
\end{equation}
and expands the scalar field about this vev, viz.
\begin{equation}
\Phi =
\left(\begin{array}{c}
\phi^+
\\
\frac{1}{\sqrt{2}} (v + \rho + i \zeta )
\end{array}\right) .
\end{equation}
The phase of $\Phi$ is then chosen using the gauge freedom 
to make $\Phi$ real.
Three components of the Higgs doublet then ``disappear''
through the gauge choice
-- the $\zeta$ and $\phi^{\pm}$ have become ``eaten'' --
to become the longitudinal components of the massive gauge bosons,
which acquire a mass term
\begin{equation}
{\cal L}_{\phi} 
=
\biggl( 1 + \frac{h}{v} \biggr)^2 
\biggl\{ m_W^2 W_{\mu}^{\dagger} W^{\mu}
+ \frac{1}{2} m_Z^2 Z_{\mu}^{\dagger} Z^{\mu} \biggr\} .
\end{equation}
Here $h=\rho$ is the remaining scalar degree of freedom
-- the Higgs boson.
The linear combination proportional to
\begin{equation}
g' W^3_{\mu} + g B_{\mu}
\end{equation}
remains massless,
\begin{equation}
m_W = m_Z \cos \theta_W = \frac{1}{2} g v
\end{equation}
and
\begin{equation}
v = ( \sqrt{2} G_F )^{-\frac{1}{2}} = 246 {\rm GeV}
\end{equation}
\begin{equation}
\lambda = \frac{m_h^2}{2 v^2} = 0.13 .
\end{equation}
The Higgs boson $h$ comes with Lagrangian terms
\begin{equation}
{\cal L}_h 
= \frac{1}{2} \partial_{\mu} h \partial^{\mu} h
- \frac{1}{2} m_h^2 h^2
- \frac{m_h^2}{2 v} h^3 
- \frac{m_h^2}{8 v^2} h^4
+ \frac{1}{8} m_h^2 v^2 .
\end{equation}

Fermion masses are constructed by contracting the Higgs doublet with the left-handed fermion doublet and then coupling to 
right-handed fermion singlets, viz.
\begin{equation}
{\cal L}_Y 
=
- y_d \bar{Q}_L \Phi d_R
- y_u \bar{Q}_L \bar{\Phi} u_R
- y_l \bar{L}_L \Phi l_R
\ + \ {\rm h.c.}
\end{equation}
which for the first generation gives
\begin{equation}
{\cal L}_Y = - \biggl( 1 + \frac{h}{v} \biggr)
\{ m_d \bar{d}d + m_u \bar{u}u + m_l \bar{l}l \} .
\end{equation}
Note that with parity violating couplings, 
the QED mass term is not possible and the 
Higgs doublet is required for fermion masses.
This means also that without the Higgs 
there is no bare fermion mass term in the Standard Model
\cite{Veltman:1997nm}.

Summarising, the fermion and gauge boson masses are then
\begin{equation}
m_f = y_f \frac{v}{\sqrt{2}} , \ \ \ \ \ 
(f = {\rm quarks \ and \ charged \ leptons})
\end{equation}
and
\begin{equation}
m_W^2 = \frac{1}{4} g^2 v^2 , 
\ \ \ 
m_Z^2 = \frac{1}{4} (g^2 + g'^2 )v^2 .
\end{equation}
The electroweak interaction Lagrangian is
\begin{eqnarray}
{\cal L} 
&=&
\bar{\Psi}_L i \gamma^{\mu} D_{\mu} \Psi_L
+
\bar{\Psi}_R 
i \gamma^{\mu} (\partial_{\mu} + i g B_{\mu}) \Psi_R
- {1 \over 4} {\rm Tr} \ W_{\mu \nu} W^{\mu \nu} 
- {1 \over 4} F_{\mu \nu} F^{\mu \nu}
\nonumber \\
& & 
+ {\cal L}_{\phi} + {\cal L}_h + {\cal L}_Y .
\end{eqnarray}
Note that gauge boson masses change 
the degrees of freedom
(with longitudinal polarisation) 
whereas fermion masses are a continuous deformation of the massless theory.
The mass dimension four Lagrangian (50) respects lepton and
baryon number conservation
although tiny effects can be induced by electroweak vacuum
tunneling processes associated with the axial anomaly \cite{tHooft:1976rip}.

The Higgs mechanism allows the gauge bosons to acquire mass
while respecting gauge invariance, which becomes ``hidden''.
Beyond the tree-level Lagrangian,
't Hooft and Veltman showed that the Higgs mechanism
is also needed to give renormalisable massive Yang-Mills
\cite{tHooft:1971qjg,tHooft:1972tcz,Veltman:1968ki}.
Further, going beyond any massive U(1) gauge bosons,
unitarity in high energy collisions involving massive 
spin-one particles 
requires that they satisfy Yang-Mills gauge invariance,
also with fermion couplings,
and that there is a Higgs providing the mass generation
~\cite{LlewellynSmith:1973yud,Bell:1973ex,Cornwall:1973tb,Cornwall:1974km}.
The Higgs propagation in intermediate states cancels 
otherwise unitarity violating terms from the longitudinal
component of the massive Z$^0$ boson.
The Higgs cannot be too heavy to do its job with maintaining
unitarity.
Indeed, if the Higgs had not been found at the LHC, 
some alternative mechanism
would have been needed in the energy range of the experiments,
e.g. 
involving strongly interacting WW scattering with the Higgs replaced by some broad resonance in the WW system 
\cite{Chanowitz:2004gk}.

\section{Renormalisation group and vacuum stability}

\begin{figure}[t!]  
\centerline
{\includegraphics[width=0.47\textwidth]
{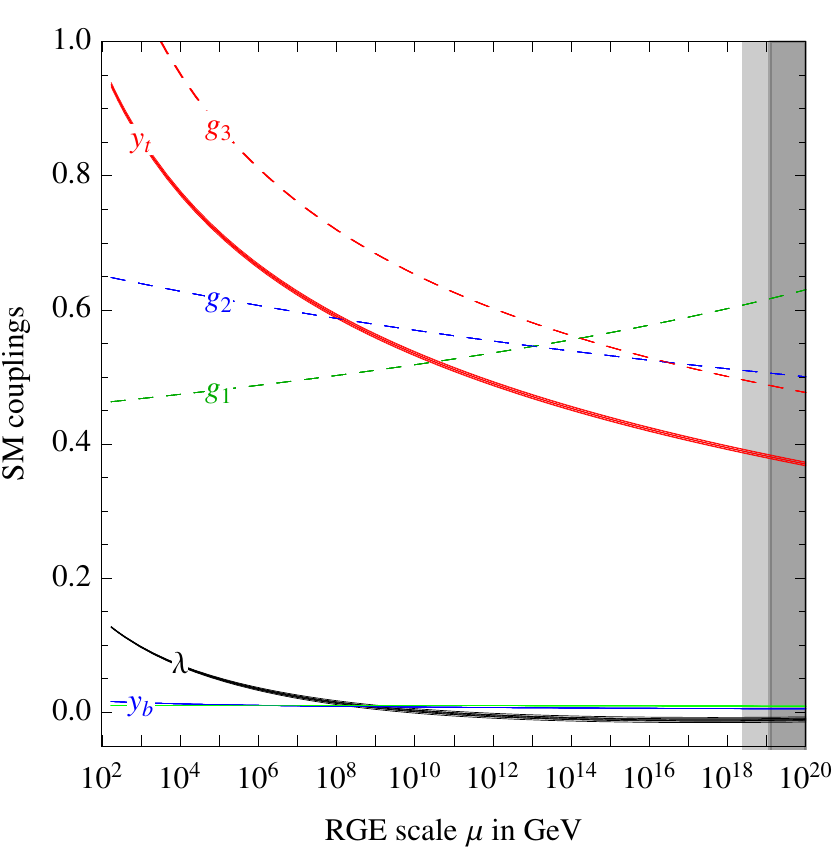}}
\caption{ 
Running couplings of the Standard Model
gauge couplings $g_1$, $g_2$, $g_3$, 
top quark Yukawa coupling $y_t$ 
and Higgs self-coupling $\lambda$.
Whether/where the Higgs self-coupling $\lambda$ 
crosses zero determines the (meta)stability of the
vacuum.
The Figure is from \cite{Degrassi:2012ry}.
}
\end{figure}

There are hints for possible critical phenomena in the ultraviolet if we can extrapolate LHC data up to close
to the Planck scale.

The scale dependence of the running couplings $\alpha_3$ 
for QCD and $\alpha_2$ for the SU(2) weak interactions 
plus the U(1) $\alpha_1$ 
are shown in Figure 1.
The non-abelian QCD and weak couplings are asymptotically 
free,
decreasing logarithmically with increasing resolution.
Their $\beta$ functions 
\begin{equation}
\beta (\alpha_i) = \mu^2 \frac{d}{d \mu^2} \alpha_i (\mu^2) 
\end{equation}
have zeros in the extreme ultraviolet at $\mu = \infty$.
In contrast, the U(1) coupling $\alpha_1$ 
increases logarithmically in the ultraviolet.
Since the Planck scale is very much less than the scale 
of any ultraviolet Landau pole, $\alpha_1$ is always finite.
With the particle masses and couplings measured at the LHC, 
the Standard Model works as a consistent theory 
with finite couplings up to the Planck scale.

The Higgs self interaction coupling $\lambda$ also runs 
under renormalisation group -- see Figure 1.
Its $\beta$-function
\begin{equation}
\beta (\lambda)  = \mu^2 \frac{d}{d \mu^2} \lambda (\mu^2)
\end{equation}
is found to have a zero close to the Planck scale.
The important issue for vacuum stability is that the 
$\beta$ function 
for the Higgs four-boson self-coupling $\lambda$ 
has a zero and when (if at all) this coupling $\lambda$ 
crosses zero.

The sign of $\beta (\lambda$) is dominated
by the large negative top quark Yukawa coupling
which yields a negative $\beta$-function at 
laboratory energies and remains negative up to (close to) 
the Planck scale.
In the absence of the large top quark coupling
(and also with the Yukawa sector switched off),
the sign of $\beta (\lambda)$ would be driven
by the positive contribution from $\lambda$ 
(and the other bosons).
It turns out that
the running of $\lambda$ is more sensitive to 
the value of 
$y_t$ than on $\lambda$ itself. 
It is a surprising property of the Standard Model and 
its specific parameter values, which leads to an
asymptotically free behavior of the Higgs boson self-coupling up to not far below the Planck scale. 
A similar flip of the renormalisation group behaviour 
applies to the top quark Yukawa coupling. 
A pure Yukawa model would
not behave as an asymptotically free theory. 
It is the interplay with QCD 
(with the top quark strongly interacting)
that yields a negative $\beta$-function at low energies 
which remains negative up to the Planck scale. 
These properties cannot be accidental, 
because if the Standard Model parameters would change only somewhat this important behaviour would be lost, 
and the emerging low energy effective theory 
would vastly differ from the Standard Model.

Electroweak vacuum stability requires that $\lambda$ remains
positive. 
Otherwise, with $\lambda$ negative definite the vacuum is unstable.
The metastable case is
that $\lambda$ goes negative and comes back positive 
with half-life of the Universe much bigger than its 
present age -- see Figure 2.
%
Vacuum metastability is a delicate issue and requires a more sophisticated form of the potential which only appears when
radiative corrections are taken into account. 
The resulting effective potential then allows for a second minimum of the potential not far below the Planck scale. Globally, the vacuum remains unstable, 
viz. 
for large Higgs field values the potential takes the form 
$\propto \lambda_{\rm eff}(\mu = \phi) \phi^4$ 
where 
$\lambda_{\rm eff} (\mu = \phi)$ 
turns negative near Planck scale and beyond.
An unstable electroweak vacuum would require some new 
additional interaction at higher scales to stabilise it.

Taking the masses and couplings measured up to LHC energies,
one finds that the electroweak vacuum resides very close to 
the border of stable and metastable 
~\cite{Jegerlehner:2013cta,Degrassi:2012ry,Buttazzo:2013uya,Bezrukov:2012sa,Bezrukov:2014ina,Alekhin:2012py,Masina:2012tz,Hamada:2012bp,Bednyakov:2015sca}
-- see Figure 3 --
suggesting possible new critical phenomena in the ultraviolet.
The Higgs vacuum sits
within 1.3 standard deviations of being stable on relating the top quark 
Monte-Carlo and pole masses if we take just 
the Standard Model interactions with no coupling 
to undiscovered new particles 
and evolve using three-loop perturbative evolution and 
two-loop matching conditions up to the highest scales of order the Planck mass \cite{Bednyakov:2015sca}. 
In these calculations electroweak vacuum stability is very sensitive to Higgs couplings,
especially the values of the Higgs and top quark masses and 
to the technical details of higher-order radiative corrections.
In calculations with a metastable vacuum $\lambda$ 
typically crosses zero around $10^{10}$ GeV
\cite{Degrassi:2012ry,Buttazzo:2013uya} 
whereas it remains positive definite with a stable vacuum
\cite{Jegerlehner:2013cta}.
For the measured value of $m_t$, $m_h$ is very close to the smallest value to give a stable vacuum. 
The 1.3$\sigma$ difference from a stable vacuum
is reduced if one includes the difference,
about 600 MeV, 
in the top quark Monte-Carlo and pole mass definitions discussed in \cite{Butenschoen:2016lpz}.
One finds a delicate interplay of Standard Model 
masses and couplings and the physics of the deep ultraviolet.
This opens the possibility that the Higgs scale might
be set by physics close to the Planck scale
with an implicit reduction 
in the number of fundamental parameters 
\cite{Jegerlehner:2013cta,Bednyakov:2015sca,Bjorken:2001yv}.
The Standard Model might behave as an effective 
theory with characteristic energy close to the Planck scale
with no new scale 
between the electroweak scale and close to the Planck mass.

\begin{figure}[t!]  
\vspace{2cm}
\centerline{\includegraphics[width=0.57\textwidth]
{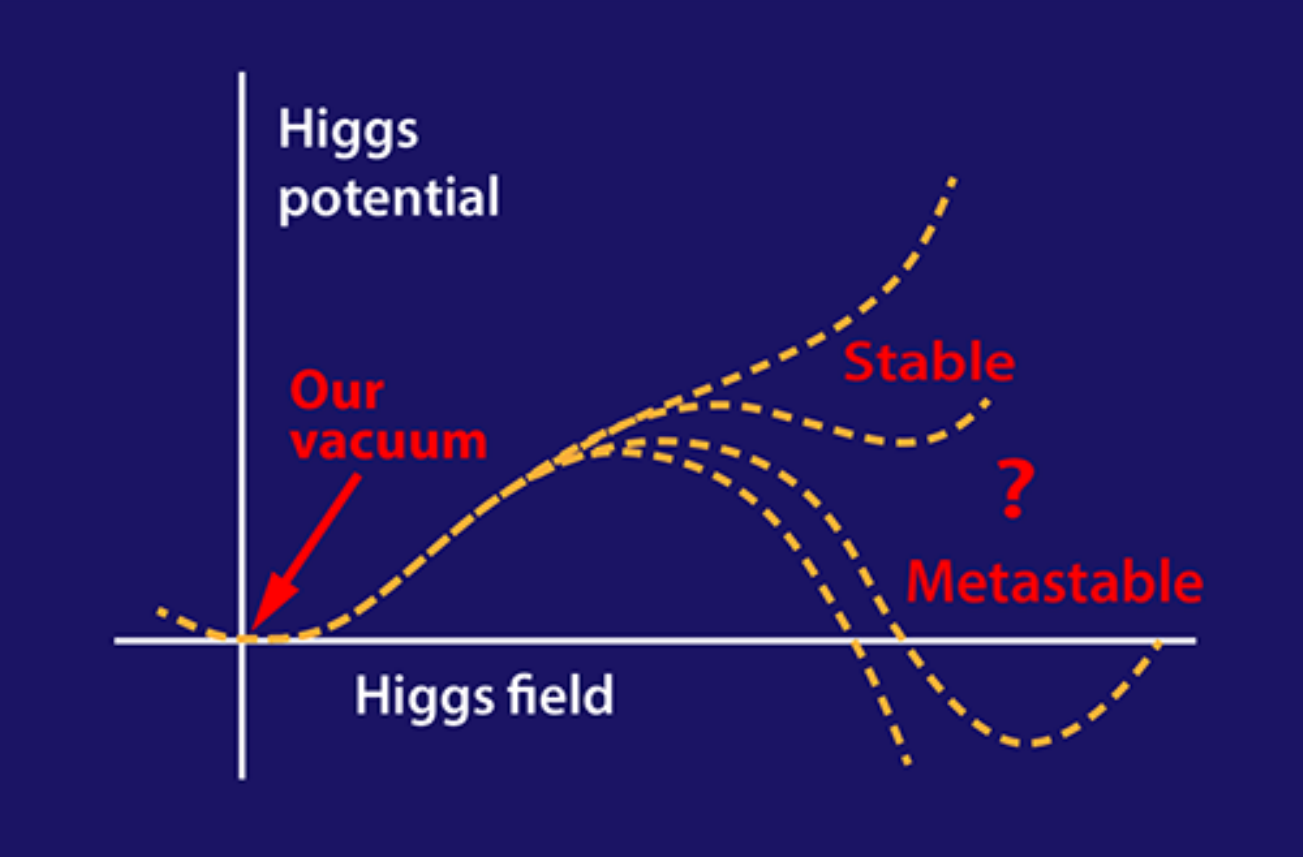}}
\caption{ 
Vacuum stability scenarios. 
The Figure is from \cite{apsphysics}, APS/Alan Stonebraker.
%
Here the curve is normalised 
with the minimum of ``our vacuum'' at 
$|\Phi| = \frac{v}{\sqrt{2}}$ with $v=246$ GeV,
see Eqs.(37) and (45).
}
\end{figure}

The Higgs vacuum sitting ``close to the edge'' of stable and metastable suggests possible 
new critical phenomena in the ultraviolet
\cite{Jegerlehner:2013cta,Degrassi:2012ry,Buttazzo:2013uya}.
One interpretation is a statistical system in 
the ultraviolet close to the Planck scale close to its
critical point.
Criticality might be an attractor point in the dynamical
evolution~\cite{Degrassi:2012ry,Buttazzo:2013uya}.
For interpretation in terms of 
multiverse ideas see~\cite{Degrassi:2012ry,Buttazzo:2013uya}.
The theoretical challenge is to identify the universality class of theories which have the Standard Model (plus any
new particle interactions waiting to be discovered) as their
long range asymptote~\cite{Jegerlehner:2013cta}.

\begin{figure}[t!]  
\centerline{\includegraphics[width=0.77\textwidth]
{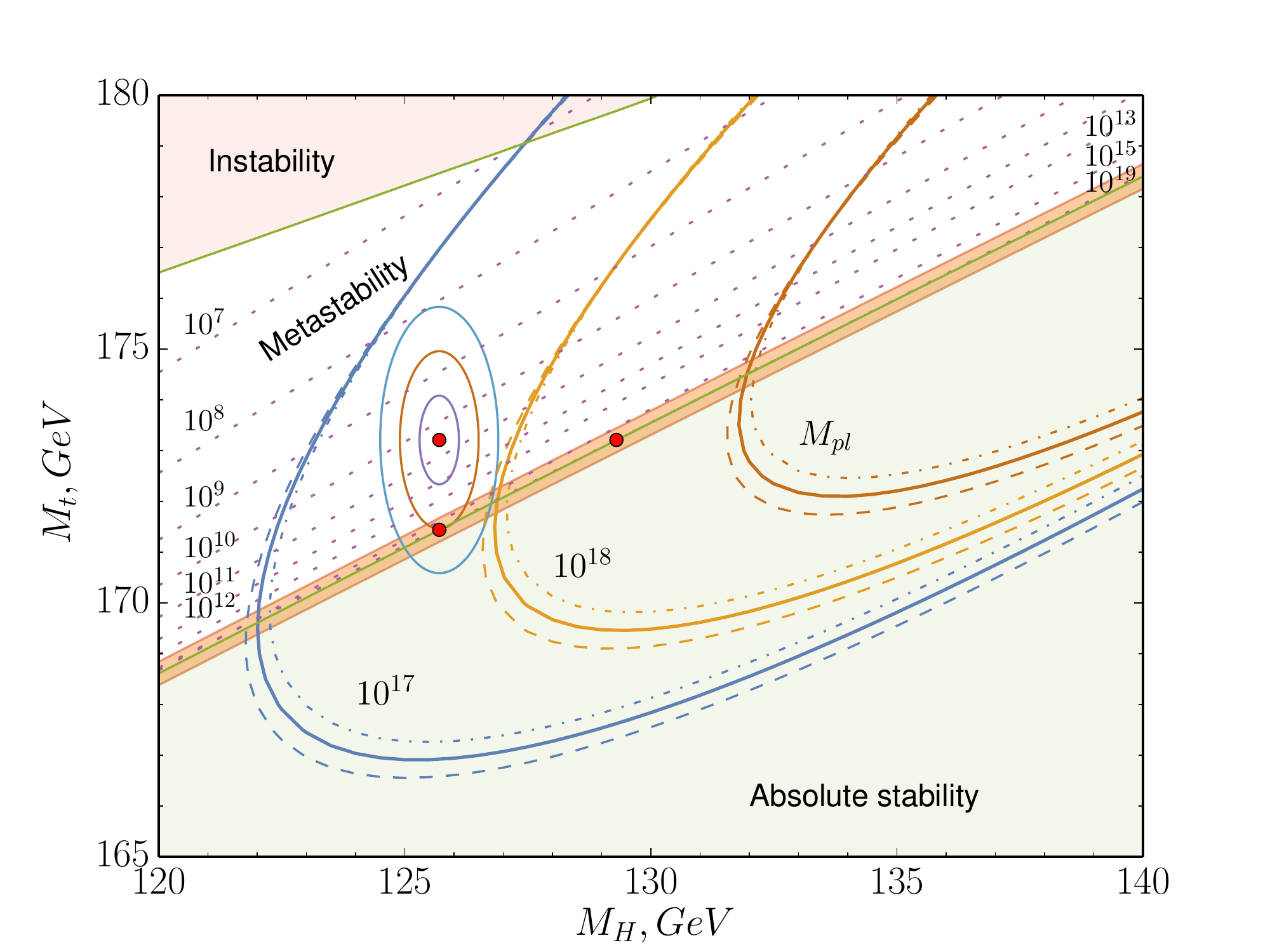}}
\caption{  
Vacuum (meta)stability of the Standard Model
showing the Monte-Carlo top quark mass and Higgs mass
with one, two and three standard deviations ellipses.
The Figure is from \cite{Bednyakov:2015sca}. 
}
\end{figure}

A possible extra source of critical phenomena in the 
ultraviolet is connected to quantum corrections to 
the Higgs mass.
The Standard Model Higgs mass squared comes with a quadratically divergent counterterm.
The renormalised mass squared is related to the bare mass term by
\begin{equation}
m_{h \ {\rm bare}}^2 
= m_{h \ {\rm ren}}^2 + \delta m_h^2
\end{equation}
where
\begin{eqnarray}
\delta m_h^2 
&=& 
\frac{\Lambda^2}{16 \pi^2}
\biggl( 
12 \lambda + \frac{3}{2} g_1^2 + \frac{9}{2} g_2^2
- 12 y_t^2 
\biggr)
\nonumber \\
&=&
\frac{\Lambda^2}{16 \pi^2}
\frac{6}{v^2} 
\biggl(
m_h^2 + m_Z^2 + 2 m_W^2 - 4 m_t^2
\biggr) .
\end{eqnarray}
Here
$\lambda$ is the Higgs self coupling, 
$g_1$ and $g_2$ are the U(1) and SU(2) couplings and
$y_t$ is the Yukawa top quark coupling to the Higgs;
$v$ is the Higgs vacuum expectation value, about 246 GeV
at the scale of the experiments.
The particle masses and Higgs Yukawa couplings are related 
through Eqs.~(48, 49).
Here we take the same ultraviolet cut-off scale $\Lambda$ 
for each particle and,
for simplicity, include just the heaviest $t, h, W, Z$ particles.
The effect of including the two-loop order correction 
is moderate as
discussed in \cite{Jones:2013aua,Jegerlehner:2013cta}.

The Higgs mass hierarchy puzzle enters when $\Lambda$ is 
taken as a physical scale.
Why is the physical Higgs mass so small compared to the 
cut-off?
For the Standard Model, the quadratic divergence in the Higgs mass self energy would cancel if the coefficient of 
$\Lambda^2$ vanishes, viz.
\begin{equation}
2 m_W^2 + m_Z^2 + m_h^2 = 4 m_t^2 .
\end{equation}
This equation is the Veltman condition \cite{Veltman:1980mj}.
The Higgs mass hierarchy puzzle would be resolved if the Veltman condition would hold
(at some scale) as a collective 
cancellation between fermion and boson contributions.
The Veltman condition does not work with 
the PDG masses
$m_t = 173$ GeV, $m_W = 80$ GeV and $m_Z = 91$ GeV,
where one would need a Higgs mass about 314 GeV 
before renormalisation group evolution.
With $m_h=125$ GeV, $\delta m_h^2$ exceeds $m_h^2$ 
for $\Lambda$ values bigger than about 600 GeV.

The terms in the coefficient of $\Lambda^2$ enter with 
different signs for boson and fermion contributions.
Further,
the running masses and couplings each have different 
renormalisation group scale dependence. 
This means there is a chance of crossing zero at some 
much higher scale.
The scale of Veltman crossing is calculation dependent.
Values reported are
$10^{16}$ GeV with a stable vacuum
\cite{Jegerlehner:2013cta}, 
about $10^{20}$ GeV
\cite{Masina:2013wja} 
and much above 
the Planck scale of $1.2 \times 10^{19}$ GeV
\cite{Degrassi:2012ry} 
with a metastable vacuum.
With the Standard Model evolution code~\cite{Kniehl:2016enc}
crossing is found at the Planck scale with a Higgs mass about
150 GeV, and not below with the measured mass of 125 GeV
\cite{sdbjk}.
These results are very sensitive to the value of the top quark Yukawa coupling, 
which presently is known only via the top quark mass and
conceptually is difficult to measure,
with connection to differences in the Monte-Carlo and pole masses.

Veltman crossing means that the renormalised and bare 
Higgs mass squared first coincide.
If we can extrapolate above this scale, 
then the bare mass squared changes sign with first order phase transition
to a symmetric phase with vanishing Higgs vev 
\cite{Jegerlehner:2013cta}.
In this case
the W and Z bosons and charged fermions become massless,
and the Nambu-Goldstone modes eaten to become 
the longitudinal components of the W and Z are liberated 
to become massive spin-zero bosons.
Jegerlehner argues that 
in this scenario the Higgs might act as 
the inflaton at higher mass scales in a 
symmetric phase characterised 
by a very large bare mass term 
for the Higgs scalars and vacuum energy contribution, 
about $10^{15}$ GeV \cite{Jegerlehner:2014mua}.
Alternatively,
the Higgs might be emergent from some more primordial 
degrees of freedom at the scale of electroweak 
symmetry breaking together with the onset of mass generation.

\section{Possible emergent gauge symmetries in particle physics}

At this point it is helpful to recollect where we are.

The tremendous success of the Standard Model at predicting and explaining the results of all our experiments, 
together with the curious result that
the Standard Model Higgs vacuum sits close to the border of
stable and metastable when LHC measured couplings are extrapolated up to the Planck scale raise interesting questions
about its symmetries.
Might the Standard Model gauge symmetries be more special
than previously expected?
Perhaps the Standard Model is an effective theory 
with characteristic energy close to the Planck scale.
Might there be some new critical phenomena in the ultraviolet
with Standard Model physics as the long range tail of a critical Planck-scale system?

\begin{table}[t!]
\centering
\caption{Typical operators in a low energy expansion.
The large ultraviolet scale $M$ is expected 
to be greater than about $10^{15}$ GeV and 
close to the Planck scale.
Table adapted from \cite{Jegerlehner:2014lba}.
}
\label{tab:LEE}
\begin{tabular}{ccccc}
\\
\hline
 & dimension & operator & scaling behaviour & \\
\hline
&&&\\[-3mm]
&$\cdot$&$\infty$--many&&\\
&$\cdot$&irrelevant &&\\
$\uparrow$&$\cdot$&operators&&\\
no&&&&\\
data & $d=6$ & $ (\Box \phi)^2, (\bar{\psi}\psi)^2, \cdots $&
 $ (E/M)^2$& \\
$|$ & $d=5$ & $ \bar{\psi}\sigma^{\mu\nu}F_{\mu\nu}\psi, \cdots $&
 $ (E/M)$ &\\
&&&&\\
\hline
&&&&\\
{\begin{tabular}{c}$|$\\
experimental \\ data \\ $\downarrow$\end{tabular}} & $d=4$ & $(\partial \phi)^2, \phi^4, (F_{\mu\nu})^2, \cdots $&
 $ \ln (E/M)$ \\
&&&&\\
\cline{2-5}
&&&&\\
 & $d=3$ & $\phi^3, \bar{\psi}\psi $&
 $ (M/E)$ & 
\\
 & $d=2$ & $ \phi^2, (A_\mu)^2 $&
 $ (M/E)^2$ & \\
 & $d=1$ & $\phi$&
 $ (M/E)^3$&\\
&&&&\\
\hline
\end{tabular}
\end{table}

With the Standard Model gauge symmetries taken as emergent 
and dynamically generated, 
the full theory is not the physics truncated 
to mass dimension four operators but also includes an infinite 
tower of higher-dimensional terms suppressed by powers of 
the large emergence scale -- see Table 1.
At low energies the physics is determined by a relatively
small number of operators with mass dimension at most four.
For these terms gauge invariance and renormalisability 
restrict the number of possible operator contributions
and strongly constrain the global symmetries of the system.
Extra symmetry breaking terms can occur in higher dimensional
operators which enter the action suppressed by powers of the
large scale of emergence,
for example with lepton and baryon number violation discussed 
in Section 6. 
These higher-dimensional terms only become active in the
particle dynamics when we are sensitive to mass and energy
scales close to the large emergence scale.
In the formal language of renormalisation group, 
the operators with mass dimension less than four 
are called relevant operators,
dimension four operators are called marginal operators, and
higher dimensional operators are called irrelevant operators.

The few relevant and marginal operators can be invariant 
under a wider range of field transformations than a generic irrelevant operator would be.
The effects of irrelevant operators are strongly
suppressed at low energies
(suppressed by powers of the large emergence scale), 
making it appear that the theory has a larger symmetry group.
Symmetry can be emergent in the low energy theory
even if it is not present in the underlying microscopic theory, e.g. associated with an infrared fixed point in 
the language of the renormalisation group.
This scenario differs from unification models 
which have maximum symmetry at the highest energies and
where symmetry breaking is generally understood 
as originating from spontaneous symmetry breaking, 
e.g. mass terms induced by the Higgs mechanism and chiral
symmetry breaking in QCD.
In a low-energy expansion any unitarity violating terms 
are suppressed by powers of the large ultraviolet scale $M$.

\section{The Standard Model in a low-energy expansion}

We have seen examples of higher-dimensional operators with 
the Pauli operator in QED
and Fermi four-fermion interaction with pre Standard Model
weak interactions.
Searches for evidence of higher mass dimension corrections
in LHC data are so far consistent with zero
meaning the ultraviolet cut-off scale is much above present
LHC energies, for recent discussion see \cite{Slade:2019bjo}.
We next discuss key examples where higher dimensional
operators may contribute to new global symmetry breaking 
beyond the Standard Model as truncated to mass-dimension four.

\subsection{Neutrino masses and mixings}

Tiny neutrino masses and neutrino flavour mixing are deduced
from solar, atmospheric and reactor neutrino disappearance experiments as well as from 
accelerator based appearance and disappearance experiments.
In addition, recent measurements by the T2K Collaboration 
in Japan are consistent with CP violation 
in the neutrino sector at
the level of two standard deviations \cite{Abe:2018wpn,Abe:2019vii}.
One finds differences between the lightest and second
lightest neutrino masses
\begin{equation}
\delta m^2 = 7.37^{+0.17}_{-0.16} \times 10^{-5} \ {\rm eV}^2
\end{equation}
and between the second lightest and heaviest neutrinos
\begin{equation}
|\Delta m^2| = 2.525^{+0.042}_{-0.030} \times 10^{-3} \ {\rm eV}^2 .
\end{equation}
Mixing angles are
\begin{equation}
\sin^2 \theta_{12} = 0.297^{+0.017}_{-0.016}, \ \ \
\sin^2 \theta_{13} = 0.0215^{+0.0007}_{-0.0007}, \ \ \
\sin^2 \theta_{23} = 0.425^{+0.021}_{-0.015}
\end{equation}
with measurement of the CP violating phase
\begin{equation}
\delta = 1.38^{+0.23}_{-0.20} \times \pi .
\end{equation}
These numbers are quoted without bias on the mixing ordering,
whether the neutrinos follow the normal 
($\nu_e < \nu_{\mu} < \nu_{\tau}$) 
or inverted 
($\nu_{\tau} < \nu_{\mu} < \nu_e$) 
ordering of masses~\cite{Capozzi:2017ipn}.

Neutrinos have no electric charge and might be either Dirac
or Majorana particles.
Majorana neutrinos have the 
property that each mass eigenstate with a given helicity
coincides with its own antiparticle with the same helicity.
Dirac neutrinos would come with the usual 
mass terms
$m_{\nu} {\bar \nu} \nu$
generated by Higgs-like Yukawa couplings.
Dirac neutrinos would imply the existence 
of (possibly sterile) right-handed neutrinos.
Majorana neutrinos come with mass generation through 
the Weinberg dimension-five operator~\cite{Weinberg:1979sa}
\begin{equation}
O_5 = \frac{(\Phi L)^T_i \lambda_{ij} (\Phi L)_j}{M}
\end{equation}  
which naturally explains the tiny neutrino masses with 
large values of the ultraviolet scale $M$, about $10^{15}$ GeV.
In Eq.~(60) $\Phi$ is the Higgs doublet, 
$L_i$ denotes the SU(2) left-handed lepton doublets 
defined in Eq.~(25), 
and $\lambda_{ij}$ is a matrix in flavour space.
Neutrinos and antineutrinos can be identified if 
lepton number is not conserved and not a good quantum number.
The Majorana mass term $\nu_L^T \nu_L$ violates lepton number conservation by two units 
(through the process $\nu \rightarrow \bar{\nu}$).
With lepton number non-conservation, Majorana neutrinos come 
with the experimental signature that they trigger 
neutrino-less double $\beta$-decays, 
with a vigorous experimental programme dedicated to search for evidence of this process.

In general Majorana neutrinos come with two extra CP mixing
angles.
On the basis of present measurements one cannot 
say anything about the possible size of these extra angles.

\subsection{Axion couplings}

The large $\eta'$ mass in QCD is induced by non-perturbative
glue, the detailed dynamics of which are still discussed. 
Independent of the detailed QCD dynamics one can construct 
low-energy effective chiral Lagrangians which include the 
effect of the QCD axial anomaly 
%
%
and use these Lagrangians to study 
low-energy processes involving the $\eta$ and $\eta'$
\cite{DiVecchia:1980yfw,Witten:1980sp,Leutwyler:1997yr}.
Define
$
U = e^{i (  \phi / F_{\pi}
                  + \sqrt{2 \over 3} \eta_0 / F_0 ) }
$
as the unitary meson matrix 
where $\phi = \ \sum \pi_a \lambda_a$ 
denotes the octet of would-be Goldstone bosons 
$\pi_a$ associated 
with spontaneous chiral symmetry breaking
with $\lambda_a$ the Gell-Mann matrices,
$\eta_0$ is 
the singlet boson and 
$F_0$ is the singlet decay constant 
(at leading order in the chiral expansion
 taken to be equal to $F_{\pi}$=92 MeV).
The gluonic mass contribution ${\tilde m}_{\eta_0}^{2}$ 
is introduced via a flavour-singlet potential involving 
the topological charge density $Q_t$ in Eq.~(22)
which is constructed 
so that the Lagrangian also reproduces the axial anomaly.
This potential reads
\begin{equation}
{1 \over 2} i Q_t {\rm Tr} \biggl[ \log U - \log U^{\dagger} \biggr]
+ {3 \over {\tilde m}_{\eta_0}^2 F_{0}^2} Q_t^2
\
\mapsto \
- {1 \over 2} {\tilde m}_{\eta_0}^2 \eta_0^2
\ \ \ \ \ 
\label{eq23}
\end{equation}
where $Q_t$ is eliminated through its equation of motion
to give the gluonic mass term for the $\eta'$.
The Lagrangian contains no kinetic energy term for $Q_t$,
meaning that the gluonic potential does not correspond
to a physical state; 
$Q_t$ is therefore distinct from mixing
with a pseudoscalar glueball state.
The $Q_t \eta_0$ coupling in Eq.(61)
describes a picture of the $\eta'$ as a mixture 
of chirality-two quark-antiquark and 
chirality-zero gluonic contributions~\cite{Bass:2018xmz}.

The non-perturbative gluonic topology which generates the 
gluonic contribution to the $\eta'$ mass 
also has the potential to 
induce strong CP violation in QCD.
One finds an extra term, $- \theta_{\rm QCD} Q_t$,
in the effective Lagrangian for axial U(1) physics
which ensures that the potential 
\begin{equation}
\frac{1}{2} i Q_t 
{\rm Tr} \biggl[ \log U - \log U^{\dagger} \biggr]
+ {3 \over {\tilde m}_{\eta_0}^2 F_{0}^2} Q_t^2
- \theta_{\rm QCD} Q_t 
\end{equation}
is invariant under global axial U(1) transformations 
with
$U \rightarrow e^{-2i \alpha} U$ 
acting on the quark fields
being compensated by 
$\theta_{\rm QCD} \to \theta_{\rm QCD} - 2 \alpha N_f$.

The term $\theta_{\rm QCD} Q_t$ is odd under CP symmetry.
If it has non-zero value, 
$\theta_{\rm QCD}$ induces a non zero neutron
electric dipole moment \cite{Crewther:1979pi}
\begin{equation}
d_n 
= 5.2 \times 10^{-16} \theta_{\rm QCD} 
\ e {\rm cm} .
\end{equation}
Experiments constrain
$|d_n| < 3.0 \times 10^{-26}e$.cm at 90\% 
confidence limit
or
$\theta_{\rm QCD} < 10^{-10}$ \cite{Afach:2015sja}.

Why is the strong CP violation parameter $\theta_{\rm QCD}$
so small?
QCD alone offers no answer to this question.
QCD symmetries allow for a possible $\theta_{\rm QCD}$ 
term but do not constrain its size.
The value of $\theta_{\rm QCD}$ is an external 
parameter in the theory just like the quark masses are.

Non-perturbative QCD arguments tell us that if the 
lightest quark had zero mass, 
then there would be no net CP violation connected
to the $\theta_{\rm QCD}$ term~\cite{Weinberg:1996kr}.
However, chiral dynamics tells us that 
the lightest up and down 
flavour quarks have small but finite masses.
In the full Standard Model 
the parameter which determines the size of strong
CP violation is
$ \Theta_{\rm QCD} 
 = \theta_{\rm QCD} + Arg \ det \ {\cal M}_q $,
where ${\cal M}_q$ is the quark mass matrix.
Possible strong CP violation then links QCD and the 
Higgs sector in the Standard Model that determines 
the quark masses.

A possible resolution of this strong CP puzzle is to postulate
the existence of a new very-light mass pseudoscalar called the axion 
\cite{Weinberg:1977ma,Wilczek:1977pj}
which couples through the Lagrangian term
\begin{equation}
{\cal L}_{a} =
- \frac{1}{2} \partial_{\mu} a \partial^{\mu} a
+ 
\biggl[ \frac{a}{M} - \Theta_{\rm QCD} \biggr]
\frac{\alpha_s}{8 \pi} G_{\mu \nu} {\tilde G}^{\mu \nu}
-
\frac{i f_\psi}{M} \partial_{\mu} a
\ \bar{\psi} \gamma_5 \gamma^{\mu} \psi - ...
\end{equation}
Here the term in $\psi$ denotes possible fermion couplings
to the axion $a$ with $f_\psi \sim {\cal O}(1)$.
The mass scale $M$ plays the role of the axion decay constant 
and sets the scale for this new physics.
The axion transforms under a new global U(1) symmetry,
called Peccei-Quinn symmetry \cite{Peccei:1977hh},
to cancel
the $\Theta_{\rm QCD}$ term, 
with strong CP violation replaced by the axion coupling 
to gluons and photons.
The axion here develops a vacuum expectation value 
with the potential minimised at
${\langle {\rm vac} | a | {\rm vac} \rangle} / M 
= \Theta_{\rm QCD}$. 
The mass of the QCD axion is given by \cite{Weinberg:1996kr}
\begin{equation}
m_a^2 = \frac{F_{\pi}^2}{M^2} 
\frac{ m_u m_d }{(m_u + m_d)^2} m_{\pi}^2 .
\end{equation}
Note that the axion coupling and mass enter at mass dimension five with the $1/M$ suppression factor.

Axions are possible dark matter candidates.
Constraints from experiments tells us that $M$ must 
be very large.
Laboratory based experiments based on the two-photon
anomalous couplings of the axion 
\cite{Ringwald:2015lqa}, 
ultracold neutron experiments to probe axion to gluon 
couplings~\cite{Abel:2017rtm},
together with astrophysics and cosmology constraints
suggest a favoured QCD axion mass 
between 
$1 \mu$eV and 3 meV 
\cite{Baudis:2018bvr,Kawasaki:2013ae},
which is the sensitivity range of the 
ADMX experiment in Seattle \cite{Rosenberg:2015kxa},
corresponding to 
$M$ between about $6 \times 10^9$ and $6 \times 10^{12}$ GeV.
The small axion interaction strength, $\sim 1/M$,
means that the small axion mass corresponds to a 
long
lifetime and stable dark matter candidate,
e.g.,
lifetime longer than about the present age of the Universe.
If the axions were too heavy 
they would carry too much energy out of supernova explosions, 
thereby observably shortening the neutrino arrival 
pulse length recorded on Earth in contradiction to
Sn 1987a data \cite{Kawasaki:2013ae}.

\subsection{Baryon number violation}

In addition to lepton number violation, 
there is also the possibility of baryon number violation
which can arise 
through the four-fermion operator \cite{Weinberg:1979sa,Wilczek:1979hc}
\begin{equation}
O_6 = \frac{1}{M^2} QQQL
\end{equation}
which enters with $1/M^2$ suppression.
Here $L$ and $Q$ are the lepton and quark doublets defined
 in Eq.~(25).
This leads to two body decays like
$p \to l^+ \pi^0$ with rate $\Gamma \propto m_p^5 / M^4$
where $m_p$ is the proton mass.
Present experimental sensitivity to such decays is about 
$10^{34}$ years \cite{Miura:2016krn}
from the SuperKamiokande experiment,
corresponding to an ultraviolet scale $M \sim 10^{15}$ GeV.

The dimension-six operator in Eq.~(66) 
becomes active at very high energies,
e.g. close to the start of the Universe, and might play an
important role in understanding the baryon asymmetry that 
we see. 
Left over from the early Universe, 
the number of baryons compared to antibaryons 
is finite and 
exceeds the number of photons in the Universe by a factor
 $\eta_B = (n_B - n_{\bar B}) / n_{\gamma} 
           \sim 6 \times 10^{-10}$ 
\cite{pdg:2018}.
Three conditions identified by Sakharov \cite{Sakharov:1967dj}
to explain the baryon asymmetry are baryon number 
non-conservation, CP violation and C violation so 
that processes occur at different rates 
for particles and antiparticles in the 
early Universe, and departure from thermal equilibrium.
Otherwise if the Universe starts with zero baryon number
it will stay with zero baryon number. 
Further higher dimensional operators,
with mass dimension 8,
that violate charge conjugation and 
time reversal invariance
while conserving parity are discussed in \cite{Mills:1967yok}.

\section{Conclusions and outlook: unification or emergence?}

Emergent gauge symmetry involves making symmetry instead of
breaking it.

Given the great success of the Standard Model at LHC collider energies and in precision experiments it is worthwhile to
re-evaluate our ideas about fundamental symmetries and their
origin in particle physics.
The Standard Model provides an excellent description of
nature up to the TeV scale with no evidence (so far) for
new particles or interactions in the energy range of our
experiments.
Further, the Standard Model works as a consistent theory
up to the Planck scale with vacuum that sits close to the border of stable and metastable.
Might the gauge symmetries that determine particle interactions be emergent from some new critical system which
exists close to the Planck scale?
Emergent gauge symmetries are observed in quantum condensed
matter systems.

With emergent gauge symmetries, the Standard Model is an
effective theory with action containing an infinite series
of higher-dimensional operators whose contributions are
suppressed by powers of the large scale of emergence.
In this scenario, 
the leading term (operators up to mass dimension four)
contributions
are renormalisable operators with greatest global 
symmetry.
Experimental constraints 
on the size of the Pauli term,
tiny neutrino masses and 
constraints on axion masses and proton decay 
suggest an ultraviolet scale $M$ greater than about 
$10^{10}$ GeV and perhaps between $10^{15}$ GeV and
the Planck scale of $1.2 \times 10^{19}$ GeV.
It is interesting that considerations of electroweak
vacuum stability suggest either a stable vacuum or 
metastable vacuum with the Higgs self-coupling crossing
zero in the same range of energy scales.
The $M$-scale suppressed higher dimensional terms only 
start to dominate the physics when we become sensitive
to scales close to $M$,
e.g. sensitive to physics processes which happened close 
to the start of the Universe.
That is, at the very highest energies the system 
becomes increasingly chaotic with maximum symmetry breaking
in contrast to unification models which exhibit the maximum
symmetry in the extreme ultraviolet.

Experimental signatures of this approach include 
lepton-number violating interactions at 
dimension-five
with Majorana neutrino masses and 
neutrino-less double $\beta$-decays
and, at dimension-six, baryon number violation which might
also induce proton decays in future precision experiments.
%
%
Small gauge groups are most likely preferred.

An emergent Standard Model gauge may help with explaining
open puzzles in particle physics.
The lepton and baryon number violating interactions in 
Eqs.~(60) and (66) 
become active at very high energies, 
e.g. typical of processes 
in the very early Universe, and
might play an important role in understanding 
the matter-antimatter asymmetry in the Universe.
Cosmology observations point to an energy budget of 
the Universe 
where just
5\% is built from Standard Model particles,
26\% involves dark matter 
(possibly made of new elementary particles) 
and 69\% is dark energy \cite{Aghanim:2018eyx}.
Dark energy 
-- 
within present experimental errors
the cosmological constant in Einstein's theory of 
General Relativity 
--
is commonly interpreted as the energy density 
of the vacuum \cite{Weinberg:1988cp,Sahni:1999gb,Bass:2011zz}.
On distance scales much larger than the 
galaxy the Universe exhibits a large distance flat geometry. 
Dark matter clumps together under normal gravitational
attraction whereas the cosmological constant 
is the same at all points in space-time and
drives the accelerating expansion of the Universe.
New axion like particles with masses and couplings suppressed
by powers of the large emergence scale might be a vital
ingredient in understanding the dark matter.
Possible consequences of emergent symmetries for cosmology 
will be discussed elsewhere~\cite{sdbjk}.

A key theoretical issue is the scale of emergence.
If the scale of Veltman crossing happens below the scale of
emergence, then the Higgs might play an essential role in
inflation.
If electroweak symmetry breaking and emergence were to
happen at the same scale, then the physics of inflation 
would involve totally new physics with different unknown 
degrees of freedom.
Electroweak physics comes with parity violating couplings
of the gauge bosons and possible Majorana neutrinos.
This prompts the question whether chirality (and neutrinos)
might play a special role in any ultraviolet critical 
phenomena leading to emergent gauge symmetry in the infrared.
Here it is interesting that the two-space-dimensional Ising
model at its critical point is equivalent to an effective theory of Majorana fermions.
One might speculate whether this result holds also in four
space-time dimensions.
The Ising model without external magnetic field has the same
equation of state as the quantum vacuum \cite{Bass:2014lja}.

\section*{Acknowledgments}

I thank F. Jegerlehner, J. Krzysiak, M. Praszalowicz and 
J. Wosiek for helpful discussions.

This article is based on lectures given at the Jagiellonian
University in Krakow, January 2019.

\end{document}